# Topological structures and phases in U(1) gauge theory [*]

Werner Kerler[a], Claudio Rebbi[b] and Andreas Weber[a]

[a] Fachbereich Physik, Universität Marburg, D-35032 Marburg, Germany
[b] Department of Physics, Boston University, Boston, MA 02215, USA

### Abstract

We show that topological properties of minimal Dirac sheets as well as of currents lines characterize the phases unambiguously. We obtain the minimal sheets reliably by a suitable simulated-annealing procedure.

## 1. Introduction

We investigate compact U(1) lattice gauge theory in 4 dimensions with the action [1]

$$S = \beta \sum_{\mu > \nu, x} (1 - \cos \Theta_{\mu\nu,x}) + \lambda \sum_{\rho, x} |M_{\rho,x}|$$

where $M_{\rho,x} = \epsilon_{\rho\sigma\mu\nu}(\bar{\Theta}_{\mu\nu,x+\sigma} - \bar{\Theta}_{\mu\nu,x})/4\pi$. The physical flux $\bar{\Theta}_{\mu\nu,x} \in [-\pi, \pi)$ is given by [2] $\Theta_{\mu\nu,x} = \bar{\Theta}_{\mu\nu,x} + 2\pi n_{\mu\nu,x}$. We consider periodic boundary conditions (apart from a final remark on our results for open boundary conditions).

The strength of the first order transition decreases with $\lambda$, the transition ultimately getting of second order [3]. We use this to set up a very efficient algorithm [3, 4] for the Monte Carlo simulations in which $\lambda$ becomes a dynamical variable.

[*]Contribution to LATTICE 94, International Symposium on Lattice Field Theory, Bielefeld, Germany, 1994. Research supported in part under DFG grants Ke 250/7-2 and 250/11-1 and under DOE grant DE-FG02-91ER40676

We argue that analyzing topological structures of the configurations (rather than only measuring global observables) is a promising tool. We apply it to monopole currents and Dirac plaquettes.

For loops the topological characterization is straightforward. For the actually occurring networks of monopole currents we have recently been able to present a mathematically sound characterization [3]. This is not only necessary for the unambiguous identification of physical features but also for the computer analysis of huge networks.

Here the analysis of configurations is extended to Dirac sheets and the fact that appropriately specified topological structures signal the phases is confirmed in more detail. In conclusion the general principle, not relying on the particular boundary conditions, is pointed out.

## 2. Dual-lattice structures

On the dual lattice for the current $J_{\rho,x} = M_{\rho,x+\rho}$ related to links the conservation law $\sum_\rho (J_{\rho,x} - J_{\rho,x-\rho}) = 0$ holds. The definition of current lines is such that for $J_{\rho,x} = 0$ there is no line on the link, for $J_{\rho,x} = \pm 1$ there is one line, and for $J_{\rho,x} = \pm 2$ there are two lines, in positive or negative direction, respectively. Networks of currents are connected sets of current lines. For a network **N** disconnected from the rest one gets the net current flow $\vec{f}$ with components $f_{\mu_3} = \sum_{x_{\mu_0} x_{\mu_1} x_{\mu_2}} J_{\mu_3,x}$ for $J_{\mu,x} \in \mathbf{N}$.

The Dirac string content of the plaquettes on the dual lattice is described by $p_{\rho\sigma,x} = -\frac{1}{2}\epsilon_{\rho\sigma\mu\nu} n_{\mu\nu,x+\rho+\sigma}$ for which the field equation $\sum_\sigma (p_{\rho\sigma,x} - p_{\rho\sigma,x-\sigma}) = J_{\rho,x}$ holds. Dirac plaquettes then are defined such that for $p_{\rho\sigma,x} = 0$ there is none, for $p_{\rho\sigma,x} = \pm 1$ there is one, and for $p_{\rho\sigma,x} = \pm 2$ there are two of them.

Dirac sheets are formed by connecting Dirac plaquettes with common edge and appropriate orientation so that $J_{\rho,x} = 0$ except at the boundaries of the sheets. Dirac sheet structures are not gauge invariant. However, they belong to equivalence classes which cannot be deformed into each other by gauge transformations. For given boundaries there are topologically distinct possibilities for the sheets. Thus the equivalence classes of sheets carry more information than the current networks.

To represent the equivalence classes of sheets by their members with minimal area we minimize the number of Dirac plaquettes by gauge transformations. We find that careful annealing is necessary to get the minimal structures reliably. For the annealing we use Metropolis sweeps for $P(\eta) \sim \exp(-\alpha\eta)$, where $\eta = \sum_{x,\mu>\nu} |n_{\mu\nu,x}|$, with random local gauge transformations keeping the angles in $[-\pi, \pi)$. In subsequent simulations we increase the parameter $\alpha$ in appropriate steps.

## 3. Topological analysis

The elements of the fundamental homotopy group $\pi_1(\mathbf{X}, b)$ of a space $\mathbf{X}$ with base point $b$ are equivalence classes of paths starting and ending at $b$ which can be deformed continuously into each other. Its generators may be obtained embedding a sufficiently dense network $\mathbf{N}$ into $\mathbf{X}$ and performing suitable transformations which preserve homotopy. We use the observation that by such a method only the generators of a subgroup are obtained if a given network $\mathbf{N}$ does not wrap around in all directions. This provides an unambiguous characterization of networks.

Choosing one vertex point of $\mathbf{N}$ to be the base point $b$ and considering all paths which start and end at $b$ we note that a mapping which shrinks one edge to zero length preserves the homotopy of all of these paths. Therefore, by a sequence of such mappings we can shift all other vertices to $b$ to get a bouquet of paths starting and ending at $b$ without changing the group content.

Describing a path by a vector which is the sum of oriented steps along the path, for $\mathbf{N}$ with $K_0$ vertices and $K_1$ edges the bouquet has $K = K_1 - K_0 + 1$ loops represented by vectors $\vec{s}_i$ with $i = 1, \ldots, K$ and j-th component $s_{ij} = w_{ij} L_j$ where $L_j$ is the lattice size. The bouquet matrix $w_{ij}$ then is to be analyzed with respect to the content of generators of $\pi_1(\mathbf{T}^4, b) = \mathbf{Z}^4$.

Current networks have the additional properties of path orientation and current conservation. The maps reducing the bouquet matrix $w_{ij}$, therefore, in addition to homotopy have to respect current conservation. This leads to a modified Gauss elimination procedure within which adding of a row to another one requires to subtract it simultaneously from a further row. In this way one arrives at the minimal form with rows $\vec{a}_1, \ldots, \vec{a}_r, \vec{t}, \vec{0}, \ldots, \vec{0}$ where $r \leq 4$. The significance of this gets obvious switching to the pair form with rows $\vec{a}_1, -\vec{a}_1, \ldots, \vec{a}_r, -\vec{a}_r, \vec{f}, \vec{0}, \ldots, \vec{0}$ which exhibits the relation to the net current flow $\vec{f}$ explicitly. It is to be noted that $\vec{f} \neq \vec{0}$ occurs only in very rare cases [3]. The number of independent pairs determines the number of nontrivial directions.

For the topological analysis of minimal Dirac sheets we use the network of plaquettes as a sufficiently dense auxiliary network. The bouquet matrix then is obtained as described before. The reduction of the bouquet matrix is simpler here because usual Gauss elimination applies, which gives the minimal form with rows $\vec{a}_1, \ldots,$ $\vec{a}_r, \vec{0}, \ldots, \vec{0}$. Now the number of independent vectors corresponds to the number of nontrivial directions.

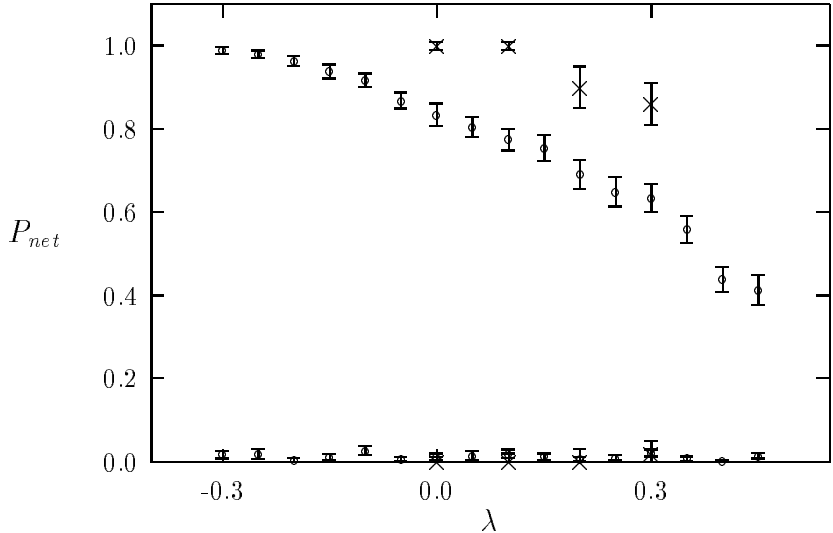

Figure 1: Probability $P_{\text{net}}$ for hot and cold phase as function of $\lambda$ for lattice sizes $8^4$ (circles) and $16^4$ (crosses).

## 4. Results

Figure 1 shows our results for the probability $P_{\text{net}}$ to find a current network which is nontrivial in four directions as function of $\lambda$ in the transition region. For larger $L$ or negative $\lambda$, where the peaks of the energy distribution related to the phases get well separated [3, 4], $P_{\text{net}}$ is seen to be 1 for the hot (confining) phase as compared to 0 for the cold (Coulomb) phase. We thus have the unambiguous topological characterization by the existence of a nontrivial network in the hot phase and its absence in the cold phase.

Our analysis of minimal Dirac sheets leads to analogous results as that of current networks. The hot phase is characterized by the existence of a topologically nontrivial sheet and the cold phase by its absence. Because, given a nontrivial current network, the sheets could be trivial or nontrivial, this answers the question about the location of the sheets.

We construct minimal Dirac sheets by making first the connections where only two Dirac plaquettes meet at an edge with appropriate orientation. Then we consider the places where more than two plaquettes meet at an edge in such a way that locally there is more than one possibility to make a connection. We find that connecting all and connecting none of the plaquettes at those places makes very little difference in the results. Thus we get the further remarkable observation that the minimal sheets do not intersect much.

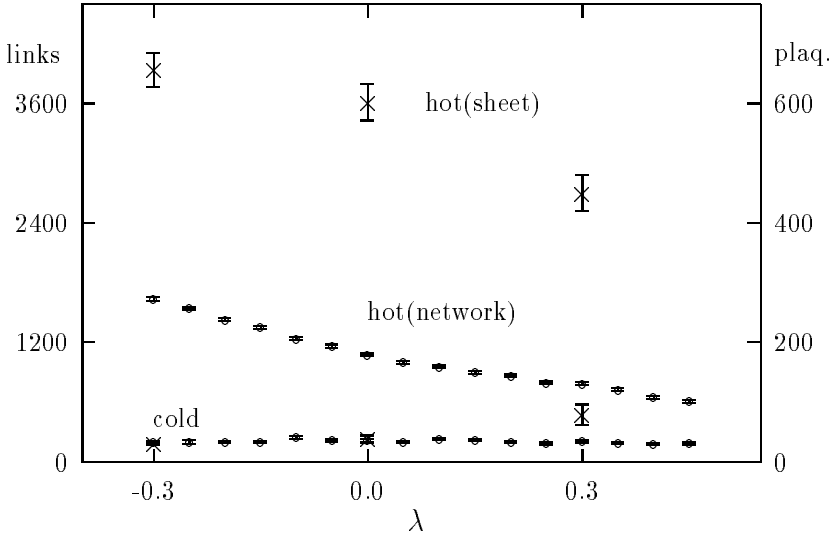

Figure 2: Size of largest current network (circles, links) and largest minimal Dirac sheet (crosses, plaquettes) in hot and cold phase as function of $\lambda$ on $8^4$ lattice.

In Figure 2 we present the sizes of the largest current network and of the largest minimal Dirac sheet. Obviously the sheets give a more sensitive signal for the phases. In Figure 3 we demonstrate that our probability $P_{\text{net}}$ is an order parameter superior to $n_{\text{max}}/n_{\text{tot}}$, the relative size of the largest network, advocated in [5].

In order to consider our results from a more general point of view we note that a topologically nontrivial structure on a finite lattice with periodic boundary conditions corresponds to an infinite structure on an infinite lattice. This suggests to characterize the hot phase more generally by the existence of an infinite network (or sheet) and the cold phase by its absence, where on finite lattices "infinite" is to be interpreted according to the particular boundary conditions. To test this characterization we have also performed simulations with open boundary conditions. They show that outside the transition region (which now is extended due to finite size effects) there is still a clear topological signal for the phases provided that in the analysis the prescription "nontrivial in all directions" is replaced by "touching the boundaries in all directions". This supports the indicated percolation-type picture.

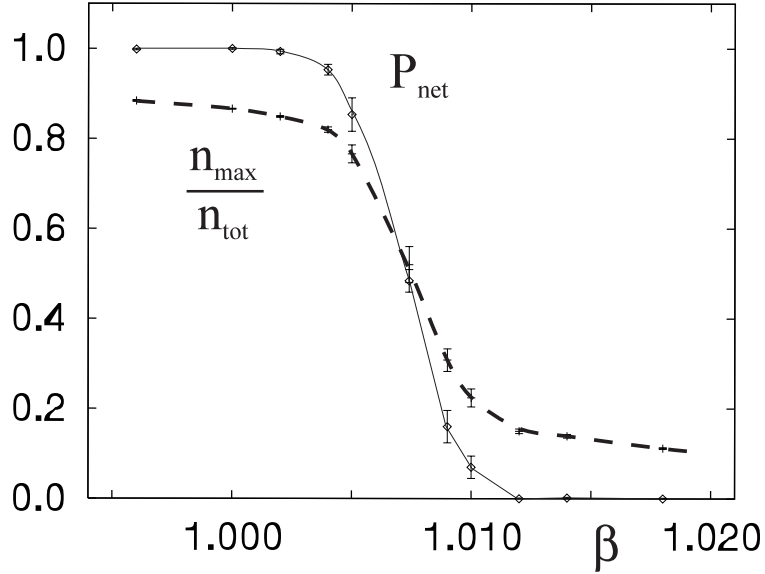

Figure 3: Order parameters $P_{net}$ and $n_{max}/n_{tot}$ for $\lambda = 0$ as function of $\beta$ on $8^4$ lattice.